\documentclass[aps, pre,twocolumn,notitlepage,longbibliography,superscriptaddress]{revtex4-2}

\usepackage[a4paper,margin=2.5cm]{geometry}
\usepackage{amsmath,amssymb,graphicx,bm}
\usepackage{hyperref}
\usepackage{xcolor}
\usepackage{physics}
\usepackage{tabularx}
\usepackage[parfill]{parskip}
\RequirePackage{color}\definecolor{RED}{rgb}{1,0,0}\definecolor{BLUE}{rgb}{0,0,1} 
\DeclareOldFontCommand{\sf}{\normalfont\sffamily}{\mathsf} 
\providecommand{\DIFaddtex}[1]{{\protect\color{blue} \sf #1}} 
\providecommand{\DIFaddbegin}{} 
\providecommand{\DIFaddend}{} 
\providecommand{\DIFdelbegin}{} 
\providecommand{\DIFdelend}{} 
\providecommand{\DIFaddFL}[1]{\DIFadd{#1}} 
\providecommand{\DIFaddbeginFL}{} 
\providecommand{\DIFaddendFL}{} 
\providecommand{\DIFdelbeginFL}{} 
\providecommand{\DIFdelendFL}{} 
\providecommand{\DIFadd}[1]{\texorpdfstring{\DIFaddtex{#1}}{#1}} 
\newcommand{\DIFscaledelfig}{0.5}
\RequirePackage{settobox} 
\RequirePackage{letltxmacro} 
\newsavebox{\DIFdelgraphicsbox} 
\newlength{\DIFdelgraphicswidth} 
\newlength{\DIFdelgraphicsheight} 
\LetLtxMacro{\DIFOincludegraphics}{\includegraphics} 
\newcommand{\DIFaddincludegraphics}[2][]{{\color{blue}\fbox{\DIFOincludegraphics[#1]{#2}}}} 
\newcommand{\DIFdelincludegraphics}[2][]{
\sbox{\DIFdelgraphicsbox}{\DIFOincludegraphics[#1]{#2}}
\settoboxwidth{\DIFdelgraphicswidth}{\DIFdelgraphicsbox} 
\settoboxtotalheight{\DIFdelgraphicsheight}{\DIFdelgraphicsbox} 
\scalebox{\DIFscaledelfig}{
\parbox[b]{\DIFdelgraphicswidth}{\usebox{\DIFdelgraphicsbox}\\[-\baselineskip] \rule{\DIFdelgraphicswidth}{0em}}\llap{\resizebox{\DIFdelgraphicswidth}{\DIFdelgraphicsheight}{
\setlength{\unitlength}{\DIFdelgraphicswidth}
\begin{picture}(1,1)
\thicklines\linethickness{2pt} 
{\color[rgb]{1,0,0}\put(0,0){\framebox(1,1){}}}
{\color[rgb]{1,0,0}\put(0,0){\line( 1,1){1}}}
{\color[rgb]{1,0,0}\put(0,1){\line(1,-1){1}}}
\end{picture}
}\hspace*{3pt}}} 
} 
\LetLtxMacro{\DIFOaddbegin}{\DIFaddbegin} 
\LetLtxMacro{\DIFOaddend}{\DIFaddend} 
\LetLtxMacro{\DIFOdelbegin}{\DIFdelbegin} 
\LetLtxMacro{\DIFOdelend}{\DIFdelend} 
\DeclareRobustCommand{\DIFaddbegin}{\DIFOaddbegin \let\includegraphics\DIFaddincludegraphics} 
\DeclareRobustCommand{\DIFaddend}{\DIFOaddend \let\includegraphics\DIFOincludegraphics} 
\DeclareRobustCommand{\DIFdelbegin}{\DIFOdelbegin \let\includegraphics\DIFdelincludegraphics} 
\DeclareRobustCommand{\DIFdelend}{\DIFOaddend \let\includegraphics\DIFOincludegraphics} 
\LetLtxMacro{\DIFOaddbeginFL}{\DIFaddbeginFL} 
\LetLtxMacro{\DIFOaddendFL}{\DIFaddendFL} 
\LetLtxMacro{\DIFOdelbeginFL}{\DIFdelbeginFL} 
\LetLtxMacro{\DIFOdelendFL}{\DIFdelendFL} 
\DeclareRobustCommand{\DIFaddbeginFL}{\DIFOaddbeginFL \let\includegraphics\DIFaddincludegraphics} 
\DeclareRobustCommand{\DIFaddendFL}{\DIFOaddendFL \let\includegraphics\DIFOincludegraphics} 
\DeclareRobustCommand{\DIFdelbeginFL}{\DIFOdelbeginFL \let\includegraphics\DIFdelincludegraphics} 
\DeclareRobustCommand{\DIFdelendFL}{\DIFOaddendFL \let\includegraphics\DIFOincludegraphics} 
\RequirePackage{listings} 
\RequirePackage{color} 
\lstdefinelanguage{DIFcode}{ 
  moredelim=[il][\color{red}\scriptsize]{\%DIF\ <\ }, 
  moredelim=[il][\color{blue}\sffamily]{\%DIF\ >\ } 
} 
\lstdefinestyle{DIFverbatimstyle}{ 
	language=DIFcode, 
	basicstyle=\ttfamily, 
	columns=fullflexible, 
	keepspaces=true 
} 
\lstnewenvironment{DIFverbatim}{\lstset{style=DIFverbatimstyle}}{} 
\lstnewenvironment{DIFverbatim*}{\lstset{style=DIFverbatimstyle,showspaces=true}}{} 

\begin{document}

\title{From Active to Odd to Smart Matter}

\author{Olivier Dauchot}
\affiliation{Laboratoire Gulliver, UMR CNRS 7083, ESPCI Paris -- PSL, 75005 Paris, France}

\date{\today}

\begin{abstract}
The study of active matter has reshaped our understanding of collective states of matter far from equilibrium by proving that energy pumped into the microscopic scale leads to order on the macroscopic scale, collective motion, and anomalous mechanical responses. More recently, the discovery of odd elasticity and nonreciprocal mechanical couplings has extended these ideas to solid-like active systems, revealing materials with nonconservative elastic response. Simultaneously, innovative developments in swarm robotics , programmable metamaterials , and learning algorithms have led to the emergence of a new frontier in which collective behavior and mechanical response are no longer fixed by design, but adapted, optimized, and learned toward functional goals.
This Perspective proposes a unifying trajectory, from active to odd to smart matter, organized along two intertwined axes: the traditional gas--liquid--solid progression of condensed matter, and the more recentparadigm shift from spontaneous collective dynamics to task-driven functionality. We try to highlight emerging principles, conceptual shifts, and open challenges that come along this trajectory, and argue that learning may play the role of a specific form of emergence, which could advantageously replace the more traditional view of control, at least in the realm of physics.
\end{abstract}

\maketitle

\section{Introduction: Beyond Equilibrium Matter}

Condensed matter has traditionally been based upon equilibrium principles -- minimizing free energy, detailed balance, reciprocal response -- to classify and predict material behavior. Active matter overcame this framework by adding systems composed of microscopic elements that persistently consume energy and perform work, inducing non-equilibrium currents, stress and flux even in steady states~\cite{Ramaswamy2010,Marchetti2013,Bechinger2016,gompper2020,te2025metareview}..
Active systems span multiple orders of magnitude in both size, energy scale and microscopic mechanism. It encompasses suspensions of bacteria and algae, cytoskeletal extracts propelled by the activity of molecular motor proteins, cell tissues, sets of self-phoretic colloidal particles, dry vibrated granular layers, or macroscopic robotic or animals swarms and flocks~\cite{Bechinger2016,gompper2020}. Yet a large fraction of these systems exhibit remarkably similar collective behaviours suggesting that activity provides a generic organizing principle, mostly agnostic to microscopic detail. And, as a matter of fact, coarse-grained descriptions confirm that collective motion, phase separation and long-range correlations can be described with some degree of universality~\cite{Toner1995,Cates2015,chate2024dynamic}.

The field is currently experiencing a second paradigm shift. In addition to knowing what collective behaviour emerges spontaneously, the focus is turning to how it can be programmed, controlled, and even learned by the material itself. This transition signals the growth of what we refer to as smart matter. The connection between active and smart matter is not merely chronological: persistent energy consumption, internal degrees of freedom, and nonequilibrium responsiveness make active systems natural substrates for adaptation and learning. We take the opportunity of this perspective to make this last statement more precise, and try to identify what are the possible consequences and promise of this shift. Our purpose is double. First we aim at clarifying the physical concepts hidden behind the new flourishing vocabulary; see also~\cite{walther2020responsive} for a complementary attempt. Second, we argue that under conditions of high dimensionality, uncertainty, distributed decision-making, and changing environments, learning may complement or sometimes outperform explicit control as a route to smart matter. In doing so, we believe that the role of physics and specifically statistical physics will be crucial.

\begin{figure*}[t]
\centering
\includegraphics[width=0.95\linewidth]{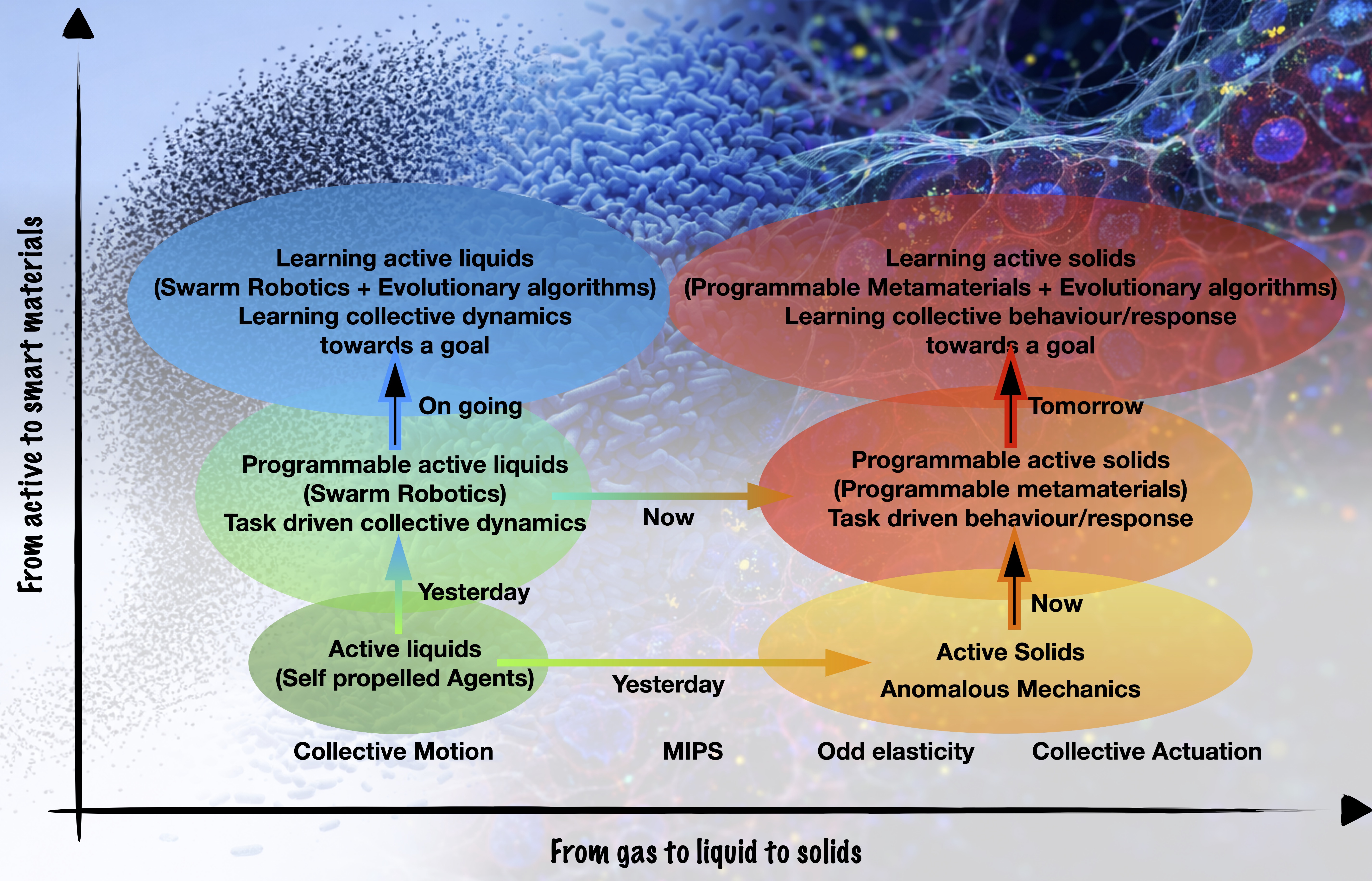}
\caption{Conceptual map from active to smart matter. The horizontal axis follows the gas--liquid--solid progression, while the vertical axis represents a transition from spontaneous collective dynamics to task-driven and learned functionality. Active liquids and solids occupy the lower-left region, while programmable and learning materials define a new regime of smart matter.}
\label{fig:map}
\end{figure*}

\section{Active Liquids: Bona Fide out of Equilibrium Phases of Matter}

Active liquids were among the first systems in which a clear demonstration was made that being driven out of equilibrium can yield bona fide phases of matter. The flocking transition~\cite{Vicsek1995,Toner1995,Chate2008,chate2024dynamic} and motility-induced phase separation (MIPS)~\cite{Tailleur2008,Cates2015} demonstrate that collective motion and phase coexistence can emerge without attractive interactions or explicit coordination. In this way, active liquids provided a well- defined setting in which the physical consequences of a steady source of microscopic energy injection could be investigated.

Active field theories, the large scale description in terms of density, polarization and stress fields provided a common framework, demonstrating how activity enters at leading order via symmetry-allowed terms that explicitly break time-reversal invariance. These theories account for instabilities, pattern formations and long-range correlations that do not occur at equilibrium ~\cite{Toner1995,Ramaswamy2010,Marchetti2013,wittkowski2014scalar,chate2024dynamic} and promoted a shift in perspective. Instead of free-energy landscapes, the phases of active liquids are governed, not only by conservation laws, but also  by dynamical symmetries and the structure of non-equilibrium currents.

Active liquids also raised important conceptual questions beyond the phenomenological realm. Foremost among them is the apparently simple question of exactly how far from equilibrium is active matter?\cite{Fodor2016}. Unlike externally driven systems, where forcing may often be treated as a perturbation, activity injects energy locally and continuously, at the scale of degrees of freedom defining the material. First attempts to map active systems onto equilibrium through e.g. effective temperatures or modified free energies only describe a limited regime, and are bound to fail in general, especially if currents, anisotropies or sustained fluxes are present \cite{Seifert2012,Fodor2016,Nardini2017}.
These difficulties motivated a sustained effort to formulate thermodynamically consistent descriptions of active matter. 
Concepts such as entropy production and non-equilibrium steady states were revisited in the context of self-propelled particles, turning active liquids into a testing ground for modern non-equilibrium statistical mechanics~\cite{markovich2021thermodynamics,johnsrud2025fluctuation}. In the process they precipitated a reevaluation of equilibrium-based intuition and inspired new quantitative characterizations for irreversibility.

Looking back, one can envision active liquids as a conceptual crucible. Their experimental richness refined theory; theoretical insights provided a language to relate systems as diverse from each other as bacterial colonies and synthetic colloids. Following the gas -- liquid -- solid sequence of condensed-material states, the obvious next step was to further increase the density and investigate active solid phases.

\section{Active Solids: From Self-Alignment to Nonreciprocal Mechanics}

\begin{figure*}[t]
\centering
\includegraphics[width=0.95\linewidth]{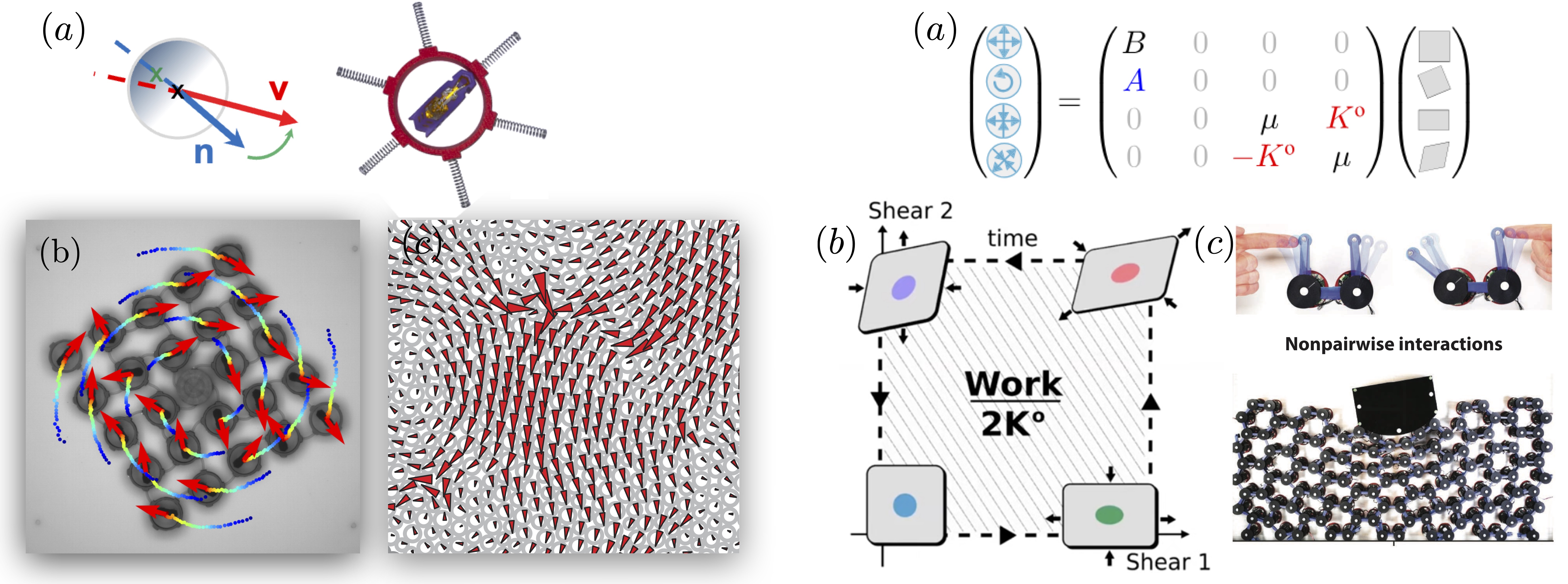}
\caption{Active Solids : nonreciprocity as a design principle (Left) (a): Self alignment, the non reciprocal coupling between orientation and displacement, leads to spontaneous collective actuation, both in (b): a model experimental system~\cite{baconnier2025self} composed of polar agents embedded inside a spring network and in (c): a minimal numerical model~\cite{Henkes2011}. (Right) Active solids with odd elasticity~\cite{Scheibner2020} is (a) characterized by the presence of additional elastic moduli $A$ and $K^o$, the presence of which allows for (b): the conversion of deformation cycles into mechanical work, and (c) nonsymmetric response, as illustrated here with a material composed of non-reciprocal units~\DIFaddbeginFL \DIFaddFL{\cite{veenstra2025adaptive}}\DIFaddendFL }
\label{fig:activesolids}
\end{figure*}

The more recent study of active solids is a major step in the evolution of active matter that takes ideas about nonequilibrium into rigid-load-bearing structures. At the conceptual level, this has occurred along two relatively independent paths that have nevertheless converged in practice. One is based on microscopic mechanisms derived and introduced for active liquids, in particular self-alignment~\cite{shimoyama1996collective,deseigne2010collective,baconnier2025self}. The other develops a more abstract approach in terms of symmetry that is based on the non-reciprocity of interactions enabled by activity \cite{Scheibner2020,Fruchart2021}.

The first path originates directly from the physics of active liquids. Self-alignment, introduced as a minimal interaction leading to flocking and collective motion, proved to be a remarkably versatile ingredient. When combined with increasing density, confinement, or explicit elastic couplings, the same microscopic mechanism offers a way to preserve non trivial collective dynamics~\cite{Henkes2011}: collective motion is replaced by collective actuation, where the dynamics condenses on a few specific non zero energy elastic modes~\cite{baconnier2022selective}. Internal degrees of freedom synchronize, leading to coherent stresses and strain and macroscopic oscillations.

The second, more abstract path to active solids builds on the idea that activity allows forces and responses that violate Newton’s third law and Onsager reciprocity, giving rise to intrinsically nonreciprocal couplings between degrees of freedom \cite{Scheibner2020,Fruchart2021}. In this picture, the nature of active solids is defined not by how activity arises microscopically but rather by how it transforms constitutive relationships. The most surprising consequence of this method is the discovery of odd elasticity: antisymmetric parts of the elastic tensor that enable a material to do work during cyclic deformation. These response coefficients are forbidden at equilibrium but become generically allowed once time-reversal symmetry is broken at the microscopic scale. This perspective makes a principle out of non-reciprocity : mechanical response is not limited by the presence of elastic potential and solids become transducers, rather than passive responders. Importantly, this description is agnostic to microscopic implementation: it applies equally to biological tissues, architected metamaterials, or abstract networks of active elements.

Although these two approaches --a microscopic, mechanism-based one and an abstract symmetry-driven one -- initially appeared distinct, they eventually meet. The key observation is that self-alignment, when embedded in a solid-like context, naturally gives rise to nonreciprocal induced-couplings among internal degree(s) of freedom. In active liquids, self-alignment translates local orientation into collective motion. In active solids, the same mechanism couples orientation, stress, and deformation in a way that breaks reciprocity. In this sense, self-alignment provides a concrete microscopic route to nonreciprocal mechanics and collective actuation. Odd elasticity at the macro-scale has not been reported yet in such active solids. Provided that nonreciprocity does not fade out with coarse-graining~\cite{dinelli2023non}, it is nevertheless expected to emerge, if not at the linear, at the nonlinear level. Seen from this perspective, collective actuation in solids is not an exotic phenomenon,  it is simply the solid-state counterpart of the same principles that give rise to collective motion in fluids.

Yet, whether liquid or solid, active matter exhibits collective states, but do not adapt them. As we argue below, this limitation sets the stage for the transition toward programmable and learning matter, where collective dynamics is no longer merely observed, but shaped toward function.

\section{Programmable Active Matter: Design, Programming, Control, and Feedback}

\begin{figure}[t]
\centering
\includegraphics[width=0.95\linewidth]{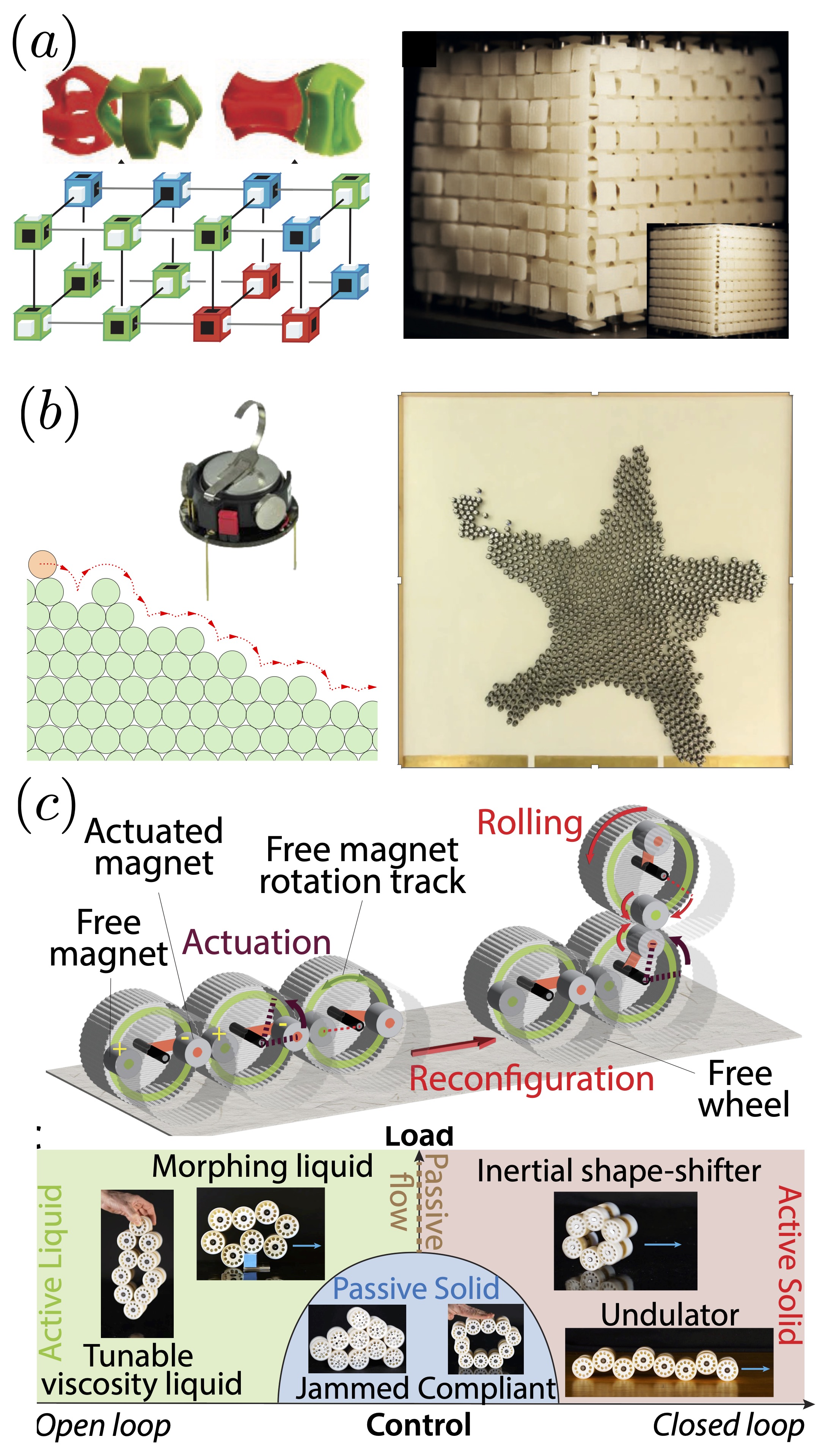}
\caption{Programmable active matter: when design turns into programs. Both traditional and more advanced metamaterials rely on externally designed interaction rules or couplings: (a) Combinatorial design of textured mechanical
metamaterials~\cite{coulais2016combinatorial}. Programmable matter allows to prescribe dynamical rules: (b) swarm of robots obey a pre-defined algorithm to assemble into a target shape~\cite{Rubenstein2014}. It also allows for control whether in open or closed loop: (c) robotic collectives with control of strength shape and motion inspired by granular media ~\cite{saintyves2024self}.}
\label{fig:program}
\end{figure}

In their journey from physics to engineering and robotics, active matter ideas kept moving the emphasis from the spontaneous emergence of collective behaviors to functionality. Local interaction rules or elastic couplings are designed so that the collective achieves a prescribed function \cite{Rubenstein2014,coulais2016combinatorial,zhang2021autonomous,bowick2022symmetry,saintyves2024self,dudek2025shape,devlin2025material,veenstra2025adaptive}. More explicitly, programmable active liquids are typically based on adaptive propulsion or interaction rules to realize various tasks such as transportation, exploration, or shape formation \cite{Rubenstein2014,zhang2021autonomous,bowick2022symmetry}. Programmable active solids exploit architected unit cells and couplings to achieve targeted mechanical responses~\cite{coulais2016combinatorial,saintyves2024self,dudek2025shape,devlin2025material,veenstra2025adaptive}.

This shift brought with it a new vocabulary -- design, program, control, feedback and learning -- that is often used interchangeably, but in fact refers to conceptually distinct approaches towards the shaping of collective behavior.  Clarifying these distinctions is important for understanding what is genuinely new in smart matter. We have tried below to stress, what we believe are the core differences between these concepts. This should by no mean be taken as a rigid classification, but a guide for physicists, who want to share a common vocabulary. In particular we have chosen to focus our distinctions on: what is fixed, what can change during operation, where information is stored, and whether past performance modifies future behavior. 

{\bf Design: from structure to function}
At the most static level, functionality can be achieved by design. In designed active matter, the intention is that microscopic rules, geometries, or couplings are fixed a priori and, hopefully, the collective exhibits the desired response. This line of practise is obviously not specific to active matter and is at the core of everyday material science. In our more specific context, examples include architected metamaterials whose unit-cell geometry encodes a target mechanical behavior, or active particles with interaction rules tuned to produce a specific pattern~\cite{coulais2016combinatorial,bowick2022symmetry}.

Design is based on a predictive understanding and not adaptation. Once fabricated, the system operates as intended -- if all goes well -- with no capability to adjust its performance in response to changing circumstances. In this respect, design extends traditional material engineering into the nonequilibrium realm without changing its basic philosophy.

{\bf Programming: prescribing rules}
In programmable active matter, the agents or elements follow conditional instructions---``if--then'' rules, state machines, Boolean rules, or lookup tables---that determine how they respond to local information.
A canonical example is that of swarm robotics: individual robots execute simple programmed rules based on their neighbors position or signals, leading to collective tasks such as aggregation, shape formation, or collective transport \cite{Rubenstein2014,Brambilla2013}. Similarly, programmable metamaterials can switch between distinct mechanical responses by toggling local constraints or activation rules~\cite{haghpanah2016programmable,ghatak2020observation}.

Programming thus differs from design in that it separates structure from function: the same physical substrate can realize multiple behaviors depending on the programmed rules. However, these rules remain externally specified and immutable during operation.

\begin{figure*}[t]
\centering
\includegraphics[width=0.99\linewidth]{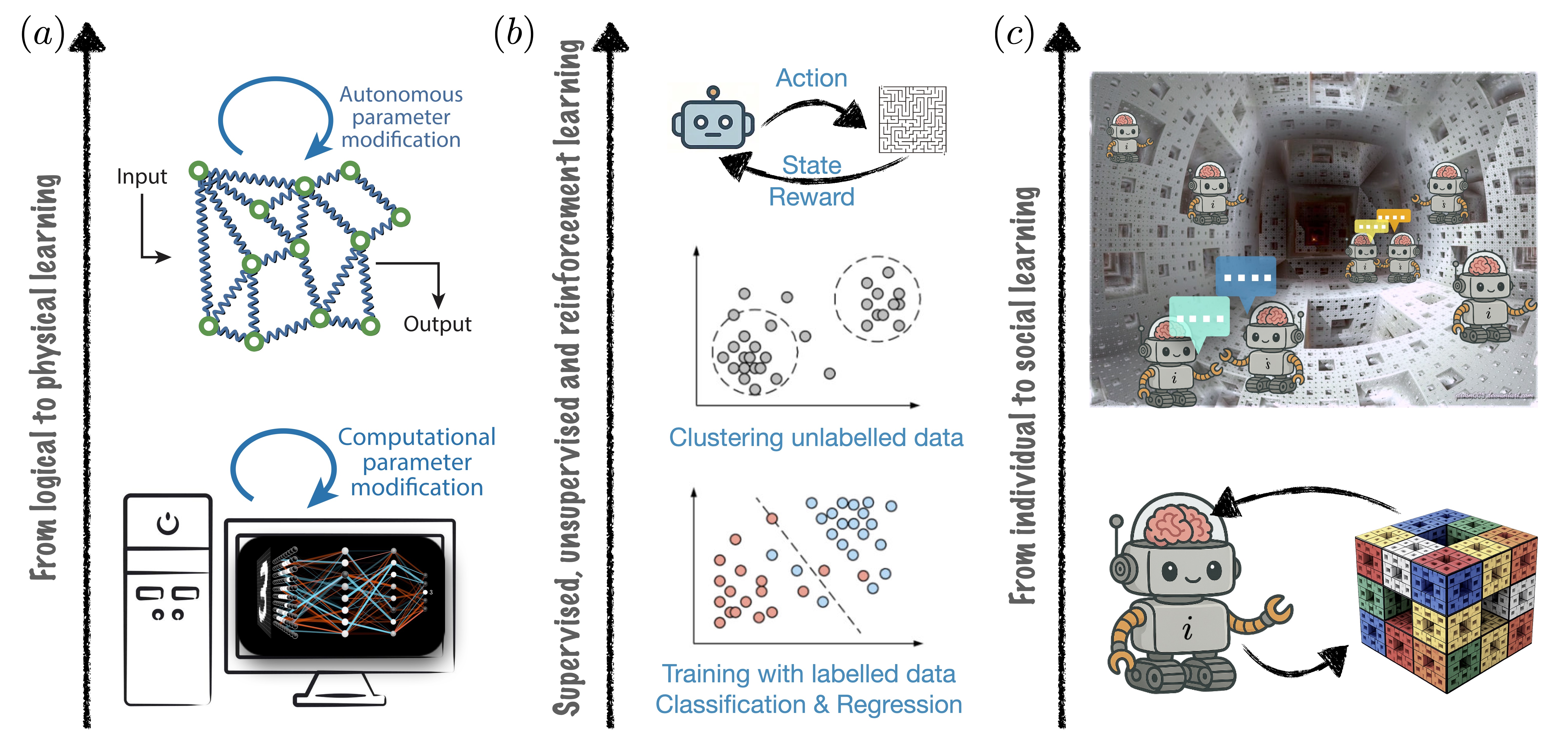}
\caption{Learning matter organized along three axes: (a)
the \emph{substrate of learning}, ranging from physical learning, where adaptation is encoded directly in material parameters or constitutive relations, to logical learning, where adaptation occurs in abstract decision rules or control policies;
(b) the \emph{mode of learning}, spanning supervised learning (explicit targets), unsupervised learning (self-organization driven by internal correlations), and reinforcement learning (optimization based on scalar reward signals);
and (c) the \emph{locus of learning}, from individual learning, where agents adapt independently, to social learning, where information is shared, propagated, or collectively selected.
Different realizations of learning matter occupy distinct regions of this space, highlighting that learning is not a single mechanism but a family of adaptive processes that reshape the equations of motion themselves.}
\label{fig:learning}
\end{figure*}

{\bf Control and Feedback: steering in real time}
Control introduces an external, or internal, entity, often called the controller,  which directly guides the operation of the system during operation using control parameters ---fields, signals, or boundary conditions--- that are adjusted in real time toward a desired state~\cite{takatori2025feedback}.
Examples include suspensions of active colloids guided by light fields, magnetic fields, or acoustic forcing, as well as robotic swarms directed by a centralized, or decentralized, controller that continuously updates global, or local, commands \cite{Bechinger2016,yan2016reconfiguring,frangipane2018dynamic,devlin2025material}. 
Control, especially decentralized control, can be very powerful, enabling rapid reconfiguration and precise steering. However, the material itself remains passive with respect to decision-making: it responds, but does not evaluate outcomes or modify its internal rules. Control does not endow the material with autonomy.

Feedback occupies a subtle intermediate position: the system outputs are measured and reinjected into the dynamics, modifying behavior based on a given rule. Feedback can stabilize collective states, enhance robustness, or correct deviations without requiring detailed microscopic control. Examples include active suspensions whose propulsion speed is adjusted based on local or global density measurements~\cite{lavergne2019group}, or robotic collectives that modify interaction strengths in response to congestion or loss of coherence~\cite{Brambilla2013,saintyves2024self}.  It is important to note that feedback does not involve learning. If the feedback law is fixed, the system reacts but does not improve. Feedback stabilizes or regulates, but it neither explores alternative strategies nor optimizes performance.

{\bf Why the distinction matters}
These distinctions highlight a fundamental boundary. Design, programming, control, and feedback rely on rules that are specified in advance, whether externally imposed or locally encoded. They shape behavior but do not generate it autonomously. The system executes a function, but does not autonomously modify its own governing rules based on accumulated performance.

This boundary becomes critical when moving toward learning matter. Learning requires not only feedback, but the ability to modify internal rules or parameters based on past performance. In this sense, learning transforms feedback from a regulatory mechanism into an exploratory one.

From a physical perspective, this marks a significant conceptual shift. Traditional active matter asks how collective behavior emerges from fixed rules. Programmable active matter asks how rules can be imposed to achieve function. Learning matter, by contrast, asks how rules themselves can emerge from interaction with the environment .

\section{Learning Matter: Physical and Logical Learning, Reinforcement, and Sociality}
Biology provides natural examples of active systems that adapt, regulate, and learn. Yet in the realm of artificial, or synthetic, materials, learning matter represents a conceptual extension of active matter. Here, collective behavior and material response are no longer fixed by design or programming, but emerge through adaptation based on experience. While the term \emph{learning} is often used loosely, it encompasses several fundamentally different mechanisms. For clarity, we distinguish three largely independent axes along which learning matter can be organized: (i) physical versus logical learning, (ii) supervised versus unsupervised and reinforcement learning, and (iii) individual versus social learning.

{\bf Physical versus logical learning}
In physical learning, adaptation is implemented directly through changes in material or dynamical parameters. Examples include elastic networks whose spring constants evolve in response to applied stresses, granular or mechanical systems that ``train'' through repeated loading cycles, or active solids that adapt their couplings to optimize a mechanical function \cite{Stern2021,stern2023learning,mandal2024learning}. In these systems, learning is embodied: memory is stored in the physical configuration or constitutive relations of the material itself.

By contrast, logical learning relies on an abstract representation of state, action, and performance, typically implemented in software or electronic control layers. Robotics provides a canonical example, where agents update decision rules or policies based on past outcomes while their physical embodiment remains unchanged. The prototypic example is that of self driving cars~\cite{badue2021self}.

On one hand, this distinction is crucial from a physics standpoint. Physical learning directly alters the equations of motion or constitutive laws, raising questions about how learning reshapes phase space, stability, and collective modes. Logical learning often treats the material as a platform rather than as the learning medium itself. On the other hand, the use of programmable matter, for which the logical units controlling the mechanical response belongs to the material, naturally blurs this boundary.

{\bf Supervised, unsupervised, and reinforcement learning}
A second axis concerns the mode of learning. The terminology of supervised, unsupervised, and reinforcement learning is inherited from computer science and robotics; the point here is not to redefine these notions, but to organize how they map onto physical material systems. In supervised learning, adaptation is guided by explicit target outputs or labeled examples. In material contexts, this may correspond to training a mechanical or active system to reproduce a prescribed response, such as a desired deformation pattern or trajectory, using externally supplied error signals~\cite{ravichandar2020recent,Stern2021}.

Reinforcement learning occupies an intermediate and particularly relevant position. Here, the system explores possible behaviors and receives scalar rewards or penalties based on performance. It is especially appealing in physical systems because it does not require detailed models or labels, only evaluative feedback. As a matter of fact the application of reinforcement learning in robotics has become a research field in its own~\cite{kober2013reinforcement}. From a physics perspective, reinforcement learning introduces an explicit or implicit optimization landscape that coexists with, may compete against, in all cases couples to the natural dynamical attractors of the system.

Finally, fully unsupervised learning relies solely on internal correlations and statistics. Examples include active systems that reorganize to enhance coherence, reduce dissipation, or maximize persistence without an externally defined goal. Such processes resonate strongly with the physical notions of self-organization and pattern selection, and thus also blur the boundary between learning and spontaneous ordering.

{\bf Individual versus social learning}
A third, often overlooked axis concerns where learning takes place. In individual learning, each agent adapts its own parameters based on personal experience. This is common in robotic agents that independently update policies or interaction strengths. 
In social learning, by contrast, information is shared, pooled, or indirectly transmitted through the collective\cite{reed2010social,dorigo1999ant,bredeche2022social}. Examples include swarms where successful strategies spread through imitation, averaging, or evolutionary selection, as well as active materials in which local adaptations propagate across the system~\cite{ben2023morphological}. Social learning introduces genuinely collective memory and raises questions familiar to statistical physics: How does information propagate? How robust is learning to noise and heterogeneity? Under what conditions does collective learning outperform individual optimization?

{\bf Why learning matter is conceptually new}
These three axes—physical versus logical learning, learning mode, and individual versus social adaptation —define a multidimensional space rather than a single category. What unifies learning matter is not the algorithmic details —many borrowed from computer science— but the fact that the rules governing collective behavior are themselves dynamical variables. In this sense, learning matter is not merely an application of machine learning to physical systems. It marks the emergence of a new class of non-equilibrium materials that do not merely move, respond, or actuate, but adapt and, over longer timescales, may even undergo evolutionary selection in the broader sense of variation and selection. From a physicist viewpoint, this represents an interesting new move away from, yet in the legacy of, traditional active matter. Activity typically breaks detailed balance; learning breaks the time translational invariance of the equations of motion. Both are made possible by the presence of internal degrees of freedom. The level of details to which we should describe them in order to capture history dependence, path selection, and functional optimization is the typical questioning statistical physics is made for.

\section{Outlook}
Active matter has been remarkably successful in revealing how simple nonequilibrium ingredients give rise to a rich zoo of collective behaviors. These behaviors are dynamical rather than structural: flocks, vortices, bands, phase-separated states, collective oscillations. Much like liquid crystalline or magnetic order in passive matter, they emerge robustly from local interactions and symmetry breaking, without the need for fine tuning.
It is therefore tempting to seek control over these collective dynamical states and turn them into functional materials, just as liquid crystalline order enabled displays and magnetic order enabled information storage. A large fraction of current efforts in programmable active matter can be read in this light, aiming at steering, stabilizing, or exploiting collective dynamics through design, programming, or control.
However controlling such large-scale dynamics in a robust and scalable way may well prove to be a chimera and the very richness that makes active matter fascinating may also render external control impractical.
More precisely, the difficulty of control is not unique. It may arise because the relevant state space is high-dimensional, because the dynamics are nonlinear, fluctuating, or sensitive to perturbations, because microscopic models are incomplete, or because the information needed to steer the system is distributed over many local degrees of freedom. Decentralized control can address part of this problem: local feedback laws and agent-level rules remove the need for a central controller and can considerably improve scalability. Yet it will face a central difficulty when the local rules themselves cannot be fixed once and for all. In that case, the problem is not merely to stabilize or steer a collective state, but to let the material modify its own rules, couplings, or policies through experience.

This observation motivates a different viewpoint. Rather than asking how to control collective dynamics, one may ask whether active systems can learn for themselves what collective behaviors are useful. 
Importantly, this does not imply that learning offers a generic plug-and-play alternative to control. It rather shifts the design problem from prescribing a complete controller to specifying an adaptive architecture: which variables may change, where memory is stored, what information is available locally or globally, and what constitutes improvement.  Different forms of learning address different limitations. Logical reinforcement learning is natural for robotic or programmable active systems that can evaluate task performance and update policies. Physical learning is better suited to materials whose couplings, geometry, or constitutive parameters can be modified directly by use. Social learning becomes relevant when information is distributed across many agents and successful strategies can spread through imitation, averaging, or selection. This last form of unsupervised learning is appropriate when there is no externally imposed target, but when the system can nonetheless improve coherence, robustness, efficiency, or persistence. In this more restricted sense, learning is expected to be useful when the system is distributed, partially observed, and exposed to changing conditions, while possessing internal degrees of freedom capable of storing the consequences of past interactions.

From this perspective, learning is not an alternative to emergence, but a new form of it: an emergent organization not only of states, but of rules. Seen this way, learning matter may be conceptually closer to the physicist’s intuition than control. Physicists are deeply familiar with emergence, self-organization, and collective selection, and perhaps less so with the centralized control of complex dynamical systems. Learning introduces history, feedback, and optimization into this familiar landscape, but does not abandon it.

Several challenges naturally emerge from this shift in perspective:
(i) How does learning interact with nonequilibrium features?
(ii) How can learning dynamics be coarse-grained and incorporated into effective field theories?
(iii) Are there universality classes of learning matter, independent of microscopic implementation?
(iv) Are there and what are the underlying organizing principle of learning matter?

In a slightly provocative formulation, the question may no longer be what materials we want to design for the future, but what futures we want materials to spontaneously design. 

\bibliography{smartmatter}

\end{document}